\documentclass[twocolumn,preprintnumbers,amsmath,amssymb,aps,prl,superscriptaddress]{revtex4}
\usepackage{graphicx}
\usepackage{textcomp} 
\usepackage{ulem} 
\usepackage[caption=false]{subfig} 
\begin{document}

\title{Cooperative Strings and Glassy Interfaces} 
\author{Thomas Salez}
\thanks{Corresponding author: thomas.salez@espci.fr}
\affiliation{Perimeter Institute for Theoretical Physics, Waterloo, Ontario, Canada}
\affiliation{Laboratoire de Physico-Chimie Th\'eorique, UMR CNRS Gulliver 7083, ESPCI ParisTech, PSL Research University, Paris, France}
\author{Justin Salez}
\affiliation{Laboratoire de Probabilit\'es et Mod\`eles Al\'eatoires, Universit\'e Paris Diderot, France}
\author{Kari Dalnoki-Veress}
\affiliation{Laboratoire de Physico-Chimie Th\'eorique, UMR CNRS Gulliver 7083, ESPCI ParisTech, PSL Research University, Paris, France}
\affiliation{Department of Physics and Astronomy, McMaster University, Hamilton, Ontario, Canada}
\author{Elie Rapha\"{e}l}
\affiliation{Laboratoire de Physico-Chimie Th\'eorique, UMR CNRS Gulliver 7083, ESPCI ParisTech, PSL Research University, Paris, France}
\author{James A. Forrest}
\thanks{Corresponding author: jforrest@perimeterinstitute.ca}
\affiliation{Perimeter Institute for Theoretical Physics, Waterloo, Ontario, Canada}
\affiliation{Department of Physics \& Astronomy, University of Waterloo, Waterloo, Ontario, Canada}

\begin{abstract}
We introduce a minimal theory of glass formation based on the ideas of molecular crowding and resultant string-like cooperative rearrangement, and address the effects of free interfaces. In the bulk case, we obtain a scaling expression for the number of particles taking part in cooperative strings, and we recover the Adam-Gibbs description of glassy dynamics. Then, by including thermal dilatation, the Vogel-Fulcher-Tammann relation is derived. Moreover, the random and string-like characters of the cooperative rearrangement allow us to predict a temperature-dependent expression for the cooperative length $\xi$ of bulk relaxation. Finally, we explore the influence of sample boundaries when the system size becomes comparable to $\xi$. The theory is in agreement with measurements of the glass-transition temperature of thin polymer films, and allows to quantify the temperature-dependent thickness $h_{\textrm{m}}$ of the interfacial mobile layer.
\end{abstract}
\maketitle

\textbf{According to Philip Anderson, the deepest and most interesting unsolved problem in solid-state physics is probably the glass transition. By extension, this includes the highly debated confinement effects in glassy polymer films. The present article introduces a minimal analytical model, that invokes only the ideas of molecular crowding and string-like cooperative rearrangement, before addressing the key effects of interfaces. The validity and simplicity of the approach make it ideal for application to various systems and geometries, and suggest that dynamics in glass-forming materials might be understood from elementary arguments.}

Glassy materials are ubiquitous in nature~\cite{Ediger2012}, and discussions about glass transition involve many areas of physics, from molecular and spin glasses to hard-sphere jamming~\cite{Gibbs1958,Mari2009,Lee2009,Krzakala2011,Bernard2011,Biroli2013}. In spite of the intense interest in the dynamical slowing that accompanies glass formation, a single microscopic theory has yet to emerge~\cite{Gotze1992,Anderson1995,Berthier2010,Liu2010,Parisi2010,Berthier2011}. Nevertheless, the phenomenological approach of free volume~\cite{Cohen1979} and the Doolittle ansatz~\cite{Doolittle1951} have been used to support the Vogel-Fulcher-Tammann (VFT) relation~\cite{Vogel1921,Fulcher1925,Tammann1926}, which describes so many of the observed behaviors. Fundamental to glass formation are the suggestions that particles are increasingly crowded, and relaxation requires the cooperative participation of a growing number of particles. The hypothesis of a cooperatively rearranging region, as introduced by Adam and Gibbs~\cite{Adam1965}, is appealing and has been observed in computational studies~\cite{Donati1998,Stevenson2006}. 

The existence of a length scale $\xi$ for cooperative rearrangement~\cite{Donth1996} has led to tremendous interest in confined glass formers, as initiated by~\cite{Jackson1990}. Perhaps, the most active example of attempts to probe $\xi$ is the study of glassy polymer films~\cite{Keddie1994,Ellison2003,Alcoutlabi2005}, where fascinating observations have been made. For the most studied case of polystyrene, reductions in the measured glass-transition temperature have been almost uniformly reported as the film thickness is reduced, both experimentally~\cite{Ediger2014} and numerically~\cite{Varnik2002}. It has been further suggested that this apparent anomaly is linked to the observed existence of a more mobile interfacial layer~\cite{Fakhraai2008,Yang2010,Chai2014,Yoon2014}. As a consequence, there have been many theoretical attempts to understand the thin-film glass transition, with varying degrees of complexity and success~\cite{Ngai1998,Long2001,Herminghaus2001,Lipson2009,Mirigian2014,Joerg2005}.

In this article, we present a simple analytical model for relaxation in glass-forming materials. First, from a microscopic molecular picture, the nature of the cooperative mechanism is explicitly defined and characterized as a function of density, and the Adam-Gibbs phenomenology is recovered. Then, by including thermal expansivity, we derive the VFT relation for the temperature dependence of the relaxation time in bulk materials. Finally, in order to address the effects of interfaces, the theory is applied to the case of thin films.

Beyond any formulation, there are two main ingredients that a microscopic cooperative theory must contain: i) ``more cooperative is easier", and ii) ``more cooperative is rarer". The first one means that, in order to redistribute a given amount of volume in a crowded environment, a cooperative rearrangement is energetically more favourable than a solitary one -- because the former is the sum of $N$ small displacements which is easier to satisfy than a single large displacement. This effect tends to maximize $N$. The second ingredient relies on the fact that glass-forming materials are made of independent particles that move incoherently due to thermal fluctuations. Therefore, it is relatively rare to have motions that are coherent in time and space and form collective objects: the larger $N$, the rarer the event. This effect tends to minimize $N$. Taken together, these two ingredients suggest a most probable value of $N$. Changing the temperature may change the crowding constraints and/or the coherence penalty, which result in the temperature dependence of this most probable value and thus the glassy behavior of interest. From numerical simulations~\cite{Donati1998,Stevenson2006,Betancourt2014} comes another important feature: it has been observed that the cooperative regions often have a fractal dimension close to $1$. This was also reported in experimental studies of repulsive colloids~\cite{Zhang2011}. Therefore, we add a third ingredient to the list above: iii) ``cooperative rearrangement is string-like". In the following, our goal is to build the simplest mean-field toy-model that contains i), ii), and iii).
\begin{figure}[t!]
\begin{center}
\includegraphics[width=8.5cm]{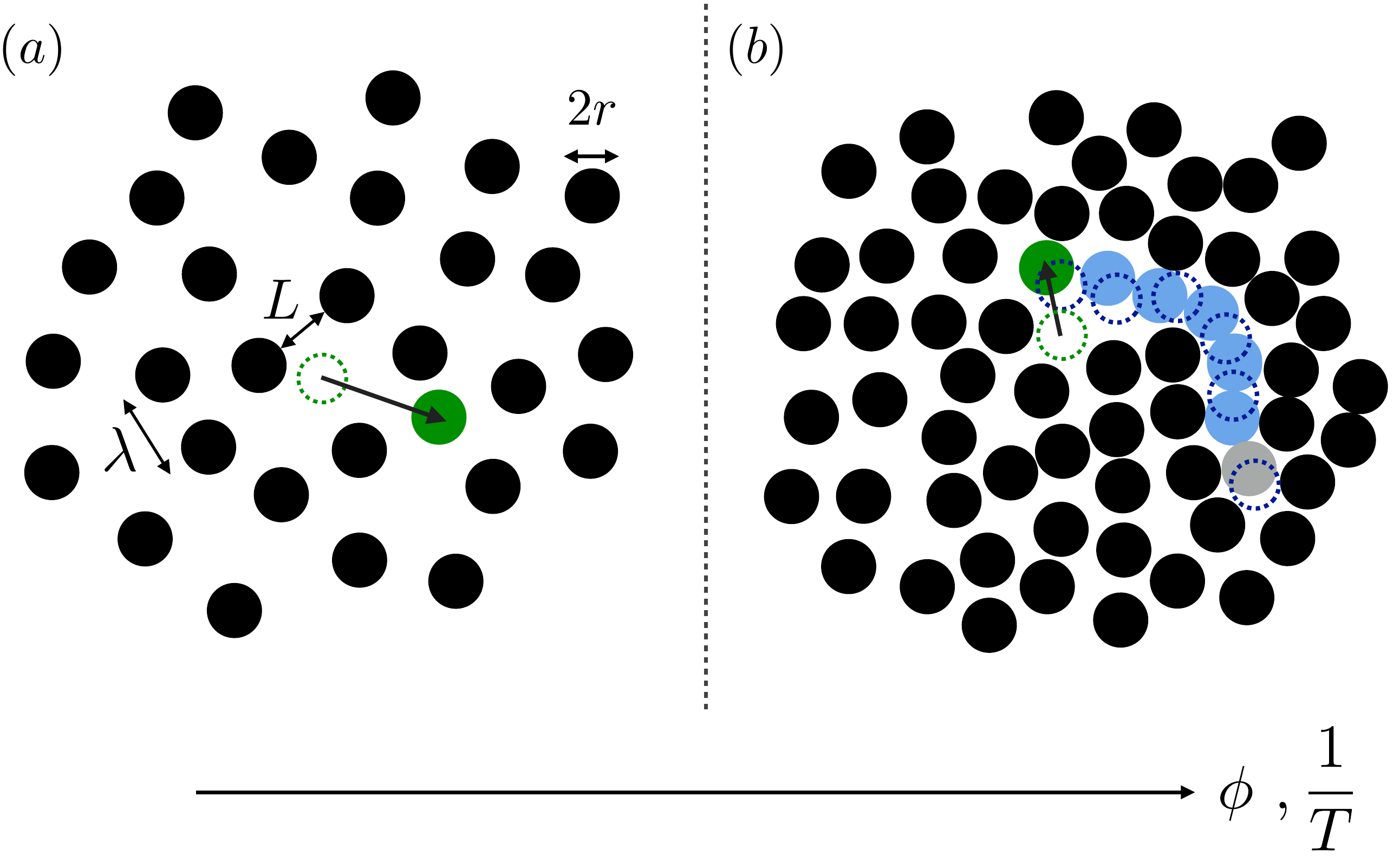}
\end{center}
\caption{\textit{Two relaxation modes in glass-forming materials. (a) At low volume fraction $\phi$, or high temperature $T$, a test particle (green) can escape its neighbouring cage by single particle motion. (b) At higher volume fraction, or lower temperature, the previous mechanism is inhibited, but relaxation of the test particle can still occur through a random string-like cooperative process: neighbours (blue) that temporarily get into close proximity with each other provide additional space for the relaxation. At the end of the string, an incoherent particle (grey) terminates the process.}}
\label{fig1}
\end{figure}

As shown in Fig.~\ref{fig1}, we consider an assembly of particles of effective radius $r$, average intermolecular distance $\lambda$, and volume fraction $\phi\propto(r/\lambda)^3$. A given particle is surrounded by a cage of nearest neighbours. Relaxation requires passage between two adjacent neighbours, the ``gate", with average size $L\sim\lambda-2r$. We define $\lambda=\lambda_{\textrm{V}}$, with $\lambda_{\textrm{V}}\sim2r$, as the point of kinetic arrest with volume fraction $\phi_{\textrm{V}}$. Note that, because we are only considering a single type of motion, and because the pressure is finite, the value of $\phi_{\textrm{V}}$ will be less than the jamming volume fraction $\phi_{\textrm{J}}$~\cite{Liu2010}. As density increases, prior to complete kinetic arrest, there exists an onset value $\phi_{\textrm{c}}<\phi_{\textrm{V}}$ of the volume fraction, and associated $\lambda_{\textrm{c}}>\lambda_{\textrm{V}}$ of the interparticle distance, at which cooperative rearrangement becomes the only possible relaxation mechanism. Note that for a cooperative motion to actually be a cooperative rearrangement, at least two particles have to exchange their positions. Note also that the value of $\lambda_{\textrm{c}}$ can be very close to $\lambda_{\textrm{V}}$ in actual physical systems, for several reasons such as deformability~\cite{Mattsson2009} and anisotropy~\cite{Donev2004} of the molecules.

As a reference, we consider the liquid-like case of a particle escaping from its cage by solitary motion (Fig.~\ref{fig1}a). On average, such motion is allowed if $\lambda>\lambda_{\textrm{c}}$, by definition of $\lambda_{\textrm{c}}$. When $\lambda\gtrsim\lambda_{\textrm{c}}$, the probability density of relaxation per unit time and per unit volume is thus $P_{\textrm{c}}\sim1/(\tau_{\textrm{c}}\,\lambda^3)$, where the constant $\tau_{\textrm{c}}$ is a typical liquid-like relaxation time at the cooperative onset. At the current level of minimal description, the Boltzmann factor associated with the sharp repulsive intermolecular potential of the gate has been replaced by an implicit Heaviside function on $\lambda-\lambda_{\textrm{c}}$. 

When $\lambda<\lambda_{\textrm{c}}$, solitary escape cannot occur since the gate is too small by an average length $\delta\sim\lambda_{\textrm{c}}-\lambda$. Relaxation is only possible through a cooperative process involving $N-1$ neighbours of the test particle, that get into close proximity with each other. Thereby, the missing space $\delta$ can be locally and temporarily made available, thus allowing for a rearrangement, as observed in bidisperse hard disks~\cite{Pal2008}. Inspired by computational studies~\cite{Donati1998,Stevenson2006,Betancourt2014} and experiments~\cite{Zhang2011}, we consider cooperative regions in the form of string-like random chains (Fig.~\ref{fig1}b). Since the gate between particles has an average length $L$, the additional length created by the cooperative move is $\Delta\sim (N-1)\,L$. The escape of the test particle through a collective motion is thus possible if $\Delta>\delta$. The threshold corresponds to $N=N^*$, where: 
\begin{equation}
\label{coopnum}
N^*(\phi)\sim \frac{\lambda_{\textrm{c}}-\lambda_{\textrm{V}}}{\lambda-\lambda_{\textrm{V}}}\sim \frac{\left(\frac{\phi_{\textrm{V}}}{\phi_{\textrm{c}}}\right)^{1/3} -1 }{\left(\frac{\phi_{\textrm{V}}}{\phi}\right)^{1/3}-1}\ .
\end{equation}
\begin{figure}[t!]
\begin{center}
\includegraphics[width=8.5cm]{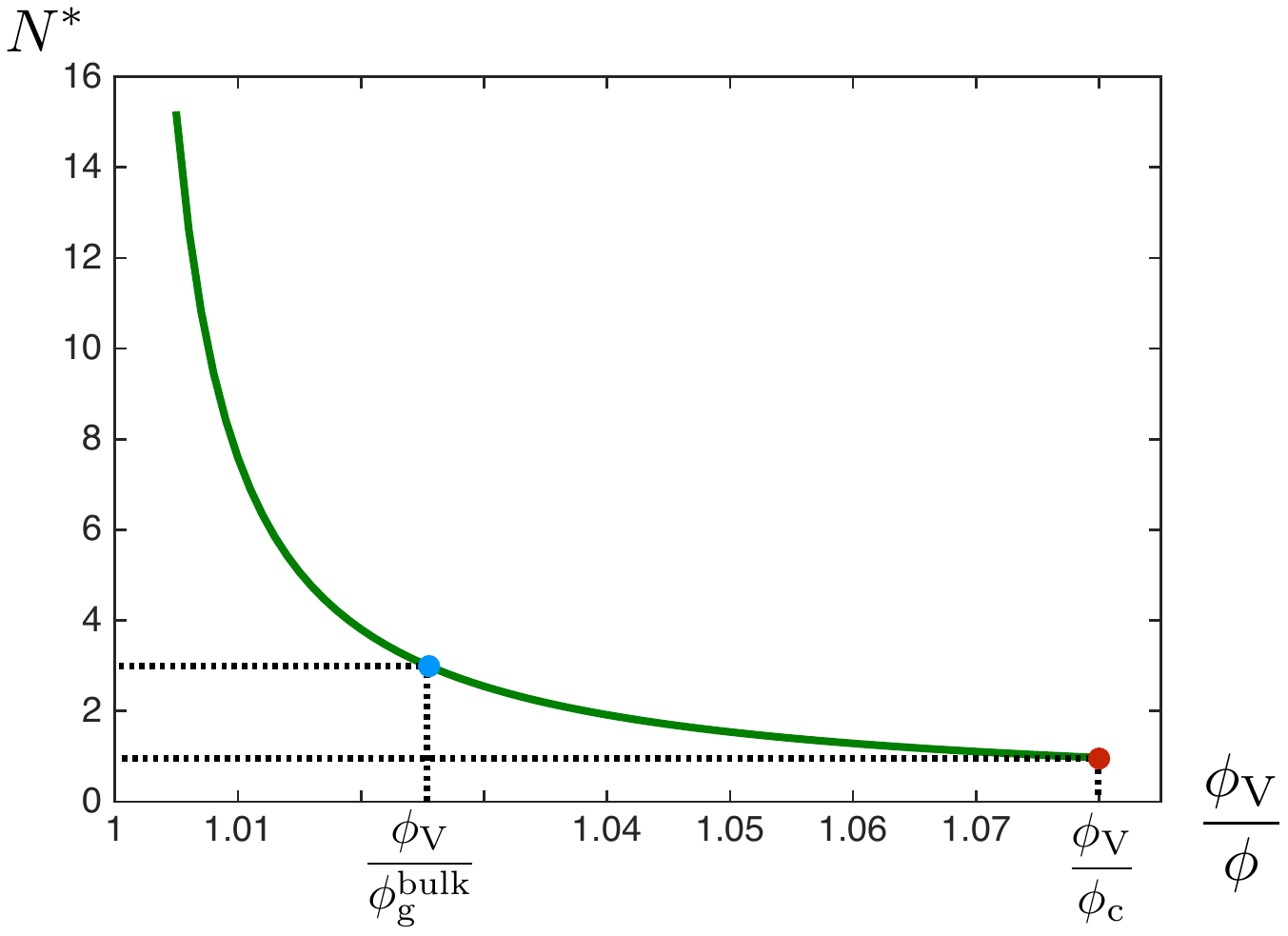}
\end{center}
\caption{\textit{Bulk cooperativity as a function of normalized inverse volume fraction (Eq.~(\ref{coopnum})). We fixed $\phi_{\textrm{V}}/\phi_{\textrm{c}}$ and $\phi_{\textrm{V}}/\phi^{\textrm{bulk}}_{\textrm{g}}$ to realistic values for polystyrene, using $\alpha\equiv-\frac{1}{\phi}\frac{\textrm{d}\phi}{\textrm{d}T}=5.5\times 10^{-4}\ \textrm{K}^{-1}$~\cite{Simona2000}, $T^{\textrm{bulk}}_{\textrm{g}} = 371\ \textrm{K}$~\cite{Rubinstein2003}, $T_{\textrm{c}}=463\ \textrm{K}$~\cite{Donth1996,Kahle1997}, and $T_{\textrm{V}}\approx322\ \textrm{K}$ (Fig.~\ref{fig4}).}}
\label{fig2}
\end{figure}
This central quantity, that we will refer to as the bulk cooperativity, is plotted in Fig.~\ref{fig2}. Note that Eq.~(\ref{coopnum}) could have been assumed, independently of any details on the string-like microscopic picture, since it is the simplest expression having the expected properties: it equals $1$ at the cooperative onset $\phi_{\textrm{c}}$; and it diverges at the kinetic arrest point $\phi_{\textrm{V}}$. The so-called bulk glass-transition point $\phi^{\textrm{bulk}}_{\textrm{g}}$ lies somewhere in between $\phi_{\textrm{c}}$ and $\phi_{\textrm{V}}$, and depends on the time scale of the observations. Naturally, such a collective scenario requires all the individual motions making up the cooperative string to be in phase. This coherence is unlikely to be spontaneously satisfied with random isotropic molecular fluctuations, resulting in a penalty. The latter can be translated into the probability factor $\sim\epsilon^{N-1}\, (1-\epsilon)$, for the independent motions of the $N-1$ consecutive particles of the cooperative string to occur coherently with the motion of the first test particle. Here, $\epsilon$ is the elementary coherence probability to be determined; and the termination factor $1-\epsilon$ expresses the incoherence of the $N+1^{\textrm{th}}$ particle (Fig.~\ref{fig1}b). Finally, the two features above can be combined to express the probability density of a cooperative relaxation process involving $N$ particles:
\begin{equation}
P_N(\phi)\sim\frac{1}{\tau_{\textrm{c}}\,\lambda^3}\,(1-\epsilon)\, \epsilon^{N-1}\ \Theta\left(N- N^*\right)\ ,
\end{equation}
where, as in the previous liquid-like case, the Boltzmann factor associated with the sharp repulsive intermolecular potential has been replaced by an Heaviside function $\Theta$. Note that $P_N(\phi)$ is maximal for $N=N^*$, at fixed $\phi$.

Summing the $P_N$ over all $N\geq N^*$, one obtains the total probability density of relaxation:
\begin{equation}
\label{main}
P(\phi)\sim P_{\textrm{c}}\ \epsilon^{\, N^*-1}\ .
\end{equation}
The relaxation is entirely determined by the cooperativity $N^*$, and is exponentially decaying with increasing $N^*$, as reported in jamming studies of air-driven granular beads~\cite{Keys2007}. Introducing the ergodic correspondence between the bulk relaxation time $\tau$ and Eq.~(\ref{main}), through $\tau P\lambda^3\sim1$, and defining the molecular time scale $\tau_0\equiv\epsilon\tau_{\textrm{c}}$, one obtains:
\begin{equation}
\label{agib}
\frac{\tau}{\tau_0}\sim\left(\frac{\tau_{\textrm{c}}}{\tau_0}\right)^{N^*}\ .
\end{equation}
This equation expresses the fact that the cooperative relaxation is a combination of $N^*$ independent motions, that are all similar to a solitary escape at the cooperative onset. Note that in a thermal description, one would estimate the onset relaxation time through the Arrhenius law: $\tau_{\textrm{c}}\sim\tau_0\,\textrm{e}^{\frac{\Delta\mu}{k_{\textrm{B}}T_{\textrm{c}}}}$, where $\Delta\mu$ is the gate energy barrier and $k_{\textrm{B}}T_{\textrm{c}}$ is the thermal energy at the cooperative onset. We thus obtain the Adam-Gibbs phenomenology~\cite{Adam1965}.

Having described the effect of crowding on the string-like cooperativity, we now study the glass transition of bulk systems. In particular, we characterize the relaxation time as a function of temperature $T$, by coupling the previous density-based picture to the thermal expansion coefficient $\alpha\equiv-\frac{1}{\phi}\frac{\textrm{d}\phi}{\textrm{d}T}$ of the equilibrium melt state. We assume that the material dilatation is small in the considered range, that goes from cooperative onset ($\phi_{\textrm{c}}$, $T_{\textrm{c}}$) to kinetic arrest ($\phi_{\textrm{V}}$, $T_{\textrm{V}}$). This is valid for several glass formers, such as polystyrene for which: $\alpha=5.5\times 10^{-4}\ \textrm{K}^{-1}$~\cite{Simona2000}, $T_{\textrm{V}}=327\ \textrm{K}$~\cite{Sahnoune1996}, and $T_{\textrm{c}}=463\ \textrm{K}$~\cite{Donth1996,Kahle1997}, so that $\phi(T)\simeq\phi_{\textrm{V}}\left[1+\alpha(T_{\textrm{V}}-T)\right]$. Combining the latter with Eqs.~(\ref{coopnum}) and~(\ref{agib}), one directly derives the VFT relation~\cite{Vogel1921,Fulcher1925,Tammann1926}, or equivalently the Williams-Landel-Ferry (WLF) relation~\cite{Williams1955}, respectively:
\begin{equation}
\label{wlf}
\tau(T)\sim\tau_0\,\exp\left(\frac{A}{T-T_{\textrm{V}}}\right)
\sim\tau_{\textrm{c}}\,\exp\left[\frac{A\ (T_{\textrm{c}}-T)}{(T-T_{\textrm{V}})(T_{\textrm{c}}-T_{\textrm{V}})}\right]\ ,
\end{equation}
where $T_{\textrm{V}}$ is identified as the Vogel temperature~\cite{Vogel1921}, and where $A\equiv(T_{\textrm{V}}-T_{\textrm{c}})\ln(\epsilon)$ is a reference temperature (see~\cite{Langer} for a thermodynamic derivation of the VFT relation based on chain-like excitations). Therefore, the elementary coherence probability $\epsilon\equiv\tau_0/\tau_{\textrm{c}}\equiv\tau_0/\tau(T_{\textrm{c}})$ is the normalized relaxation rate at the cooperative onset. The bulk glass-transition temperature $T^{\textrm{bulk}}_{\textrm{g}}$ is defined as the one at which $\tau$ reaches a given large experimental time scale $\tau^{\textrm{bulk}}_{\textrm{g}}\equiv\tau(T^{\textrm{bulk}}_{\textrm{g}})$. 

Since our model leads to the VFT and WLF time-temperature superpositions, it captures well the so-called fragile-glass phenomenology~\cite{Angell1995}, and links thermal expansion and fragility as observed in metallic glasses~\cite{Guo2007}. Strong-glass phenomenology can be recovered as well when $T_{\textrm{V}}\ll T^{\textrm{bulk}}_{\textrm{g}}$. This is reminiscent of soft colloidal glasses~\cite{Mattsson2009}, for which the kinetic arrest point is shifted to higher volume fractions due to particle deformability. Note that additional molecular processes, beyond the scope of this work, may lead to bulk relaxation that does not diverge at $T_{\textrm{V}}$~\cite{Edwards1986,Zhao2013}. Besides, as the system approaches $T_{\textrm{V}}$, the fractal dimension of the cooperative regions may change~\cite{Stevenson2006}. Finally, our cooperative-string model is a free-volume approach and, as such, is subject to the same criticism as all free-volume models~\cite{Colucci1997,Dyre2006}. 

The bulk relaxation process presented above consists of random cooperative strings and is entirely determined by $N^*$. Therefore, in the vicinity of the kinetic arrest point, the length scale $\xi$ of the cooperative regions reads $\xi \sim \lambda N^{*{\nu}}$, where: $\nu=1/2$ for a simple random walk; $\nu\approx0.588$ for a self-avoiding walk; and $\nu=1/3$ for a 3D compact shape. Consistent with our minimal string-like model, we consider the simple random walk. Invoking Eq.~(\ref{coopnum}) and thermal expansivity, one obtains:
\begin{equation}
\label{coopleng}
\xi(T)\sim\lambda_{\textrm{V}}\ \sqrt{\frac{T_{\textrm{c}}-T_{\textrm{V}}}{T-T_{\textrm{V}}}}\ .
\end{equation}
At the cooperative onset, this asymptotic expression of the cooperative length corresponds to the effective particle size $\lambda_{\textrm{V}}\sim2r$. At the Vogel temperature, $\xi$ diverges with exponent $-1/2$. Note that this power-law is similar to the one measured in 2D vibrated granular media around jamming~\cite{Lechenault2008}. The exponent also compares well to the values of $-0.59$ from kinetically constrained lattice-gas simulations~\cite{Berthier2003}, and $-0.62$ from numerical studies of Lennard-Jones liquids~\cite{Donati1999}. Unfortunately, the direct measurement of $\xi(T)$ below $T^{\textrm{bulk}}_{\textrm{g}}$ is challenging for 3D thermal glasses. A natural alternative way is thus to reduce the sample size towards $\xi$~\cite{Jackson1990,Keddie1994,Alcoutlabi2005}.
\begin{figure}[t!]
\begin{center}
\includegraphics[width=8.5cm]{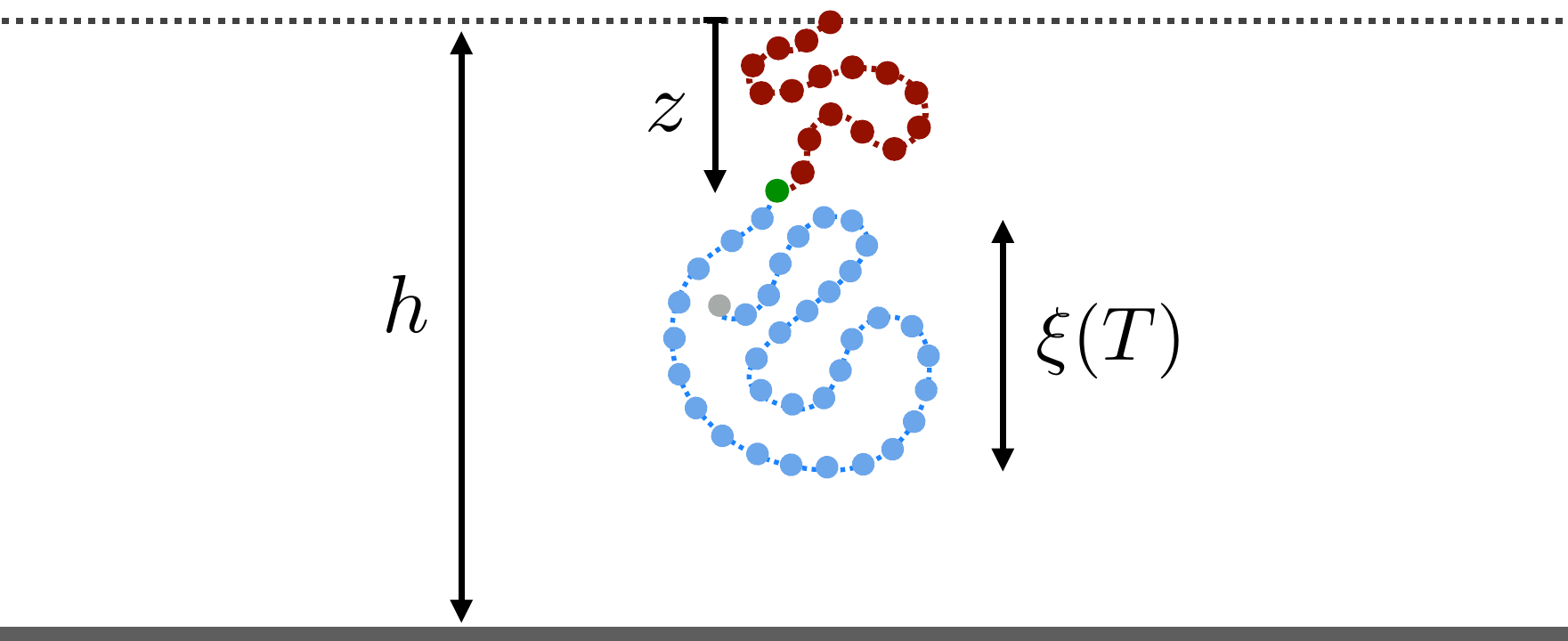}
\end{center}
\caption{\textit{Two string-like cooperative paths in a thin film of thickness $h$. For an inert supporting substrate (thick line), relaxation of a test particle (green) at a distance $z$ from a free interface (dashed line) can occur through either a bulk cooperative string (blue) of size $\xi$ (Eq.~(\ref{coopleng})), or a truncated string (red) touching the interface.}}
\label{fig3}
\end{figure}

We now turn to the case of a thin film of thickness $h$, supported on an inert substrate. Our model predicts that the key effect of the free interface is to favour a higher surface mobility, as observed in experiments~\cite{Fakhraai2008,Yang2010,Ediger2014,Chai2014,Yoon2014}. Indeed, the cooperative strings can now be truncated by the cage-free boundary (Fig.~\ref{fig3}), leading to a lower surface cooperativity. Therefore, we introduce the average local cooperativity $N_{\textrm{s}}^{\,*}(z,T)$, at a distance $z$ from the free interface. The natural length scale of the problem is the bulk cooperative length $\xi$. When $z\gg\xi$, the free interface is typically not reached with less than $N^*$ cooperative particles, and the relaxation is bulk-like with $N_{\textrm{s}}^{\,*}\sim N^*$. As $z\rightarrow0$, $ N_{\textrm{s}}^{\,*}$ vanishes due to the absence of caging at the boundary. In between, $N_{\textrm{s}}^{\,*}$ takes all the intermediate values. Thus, $N_{\textrm{s}}^{\,*}$ varies along $z$ over the bulk cooperative length $\xi$, and is expected to have the following asymptotic self-similar form in the vicinity of $T_{\textrm{V}}$:
\begin{equation}
\label{locN}
N_{\textrm{s}}^{\,*}(z,T)\sim N^*\ f\left(\frac{z}{\xi}\right)\ ,
\end{equation}
where $f$ is a continuous and monotonic function satisfying $f(0)\sim0$, and $f(u\gg 1)\sim1$. 

Since our description of the cooperative process involves random strings of particles, Eq.~(\ref{locN}) can be supported at large $N^*$ by the following argument based on Brownian motion. For a given cooperative string, we define $n_0\leq\infty$ as the number of particles at which the string hits the free interface for the first time. If $n_0\geq N^*$, the behavior is bulk-like; if $n_0< N^*$, the string is truncated by the interface. Therefore, the quantity of interest is the density probability $g$ of first passage at the interface with $n_0$ cooperative particles. Defining the first-passage ``time" $t_0=n_0/N^*$, and starting at ``distance" $Z=z/\xi$ from the interface, one gets the 1D expression: $g(t_0,Z)=(2\pi)^{-1/2}t_0^{\,-3/2}Z\exp\left[-Z^2/(2t_0)\right]$~\cite{Karatzas1988}. The local cooperativity being the minimum between $n_0$ and $N^*$, the average local cooperativity is defined as $N_{\textrm{s}}^{\,*}=N^*\left\langle\min(1,t_0)\right\rangle_{t_0}=N^*\left[1-\int_0^1dt_0\,(1-t_0)\,g(t_0,Z)\right]$, where we explicitly see the interfacial lowering of the bulk cooperativity. Calculating the integral, we recover Eq.~(\ref{locN}) with $f\left(u\sqrt{2}\right)=\textrm{erf}(u)+2u\exp\left(-u^2\right)/\sqrt{\pi}-2u^2\ \textrm{erfc}(u)$. Note that the exact functional form chosen for $f$ is not crucial when comparing to the experimental data below, as other sufficiently sharp functions provide similar results.

Following the derivation of Eq.~(\ref{agib}) and assuming that one can replace $N^*$ by the local average cooperativity $N_{\textrm{s}}^{\,*}$ of Eq.~(\ref{locN}), one obtains the local relaxation time:
\begin{equation}
\label{main2}
\frac{\tau_{\textrm{s}}(z,T)}{\tau_0}\sim\left(\frac{\tau}{\tau_0}\right)^{f\left(\frac{z}{\xi}\right)}\ .
\end{equation}
We thus see that $f$ acts as a local exponent, ranging from 0 to 1, on the normalized bulk relaxation time. This formula generalizes the Adam-Gibbs phenomenology~\cite{Adam1965} by accounting for the effect of a free interface.

\begin{figure}[t!]
\begin{center}
\includegraphics[width=8.5cm]{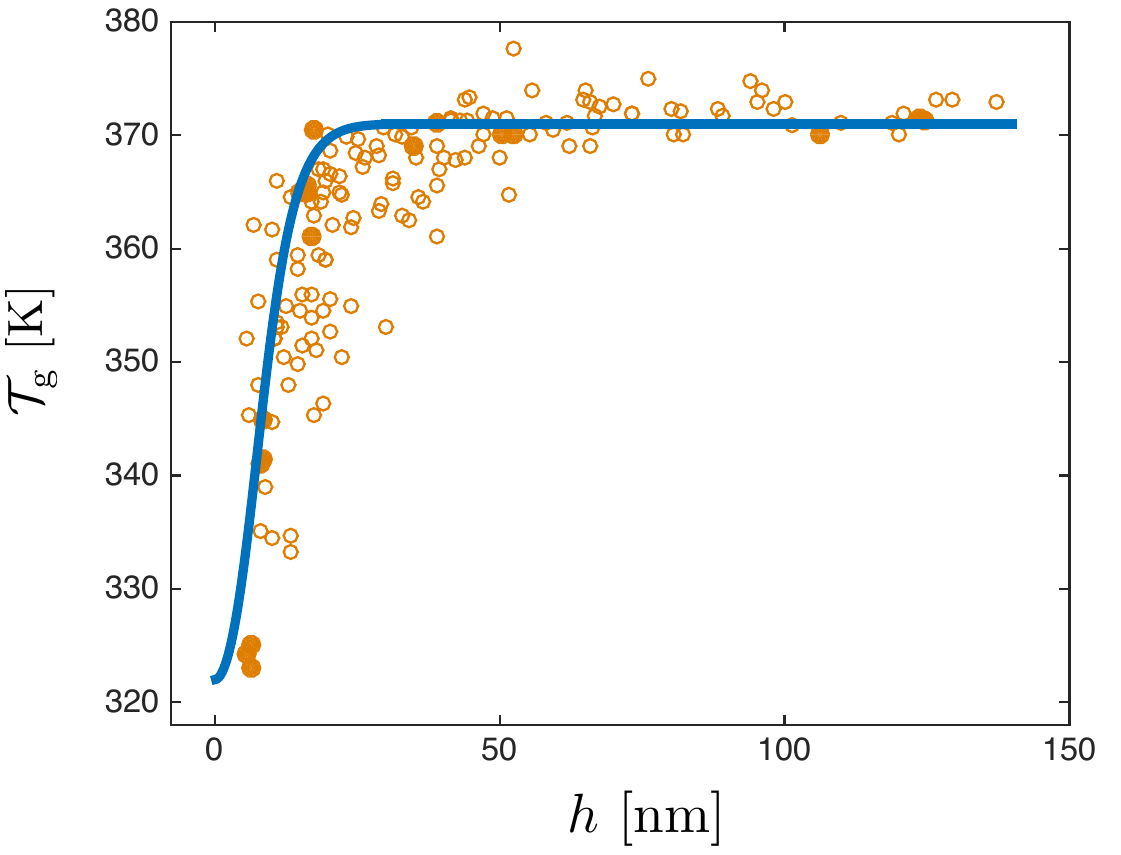}
\end{center}
\caption{\textit{Comparison between dilatometric experimental data (filled symbols) for the reduced glass-transition temperature $\mathcal T_{\textrm{g}}(h)$ of thin polystyrene films supported on silicon substrates~\cite{Raegen2008}, and the theory (line) given by Eq.~(\ref{tgh}). Other literature data~\cite{Roth2005} is shown (open symbols) for completeness. The fixed parameters are the bulk glass-transition temperature $T^{\textrm{bulk}}_{\textrm{g}} = 371\ \textrm{K}$~\cite{Rubinstein2003}, and the onset temperature $T_{\textrm{c}}=463\ \textrm{K}$~\cite{Donth1996,Kahle1997}. The two adjustable parameters are the critical interparticle distance $\lambda_{\textrm{V}}\approx 3.7$~nm, and the Vogel temperature $T_{\textrm{V}} \approx 322\ \textrm{K}$.}}
\label{fig4}
\end{figure}

Finally, we compare our theory to dilatometric measurements of reduced glass-transition temperatures in thin polystyrene films supported on silicon substrates. In the experiments, the thickness-dependent glass-transition temperature $\mathcal{T}_{\textrm{g}}(h)$ is defined as the location of a kink in the dilatation plot obtained by ellipsometry~\cite{Keddie1994}. This change of expansivity occurs when the system is half-glassy and half-liquid. Given that $f$ is monotonic, this translates to the apparent transition occurring when, at the middle $z=h/2$ of the film, the local relaxation time $\tau_{\textrm{s}}$ equals the bulk relaxation time at the bulk glass transition $\tau^{\textrm{bulk}}_{\textrm{g}}$~\cite{Forrest2014}. Invoking Eqs.~(\ref{wlf}) and~(\ref{main2}), $\mathcal{T}_{\textrm{g}}(h)$ thus satisfies:
\begin{equation}
\label{tgh}
2\ \xi\left(\mathcal{T}_{\textrm{g}}\right)\ f^{\,-1}\left(\frac{\mathcal{T}_{\textrm{g}}-T_{\textrm{V}}}{T^{\textrm{bulk}}_{\textrm{g}}-T_{\textrm{V}}}\right)=h \ ,
\end{equation}
where $f^{\,-1}$ denotes the inverse of the bijective function $f$. The solution of this equation is plotted in Fig.~\ref{fig4}, and compared to measurements on polystyrene~\cite{Roth2005,Raegen2008}. The literature data encompasses a wide variety of techniques and protocols~\cite{Roth2005}, and for purposes of fitting we consider the restricted data of~\cite{Raegen2008}, where the annealing conditions and atmosphere have been carefully controlled and documented. We see from Fig.~\ref{fig4} that our model provides excellent agreement with the experiments. The two adjustable parameters are: the critical interparticle distance $\lambda_{\textrm{V}}\approx 3.7$~nm which is reasonably found to be comparable to the persistent length of polystyrene~\cite{Wu1999,Rubinstein2003}; and the Vogel temperature $T_{\textrm{V}} \approx 322\ \textrm{K}$, which is close to the reported value of 327 K~\cite{Sahnoune1996}. Note that we have only considered polystyrene, as that material is extremely well studied. However, by varying the single parameter $\lambda_{\textrm{V}}$ in Eqs.~(\ref{coopleng}) and~(\ref{tgh}), one would obtain larger, smaller, or even immeasurable reductions in $\mathcal{T}_{\textrm{g}}$, as could be observed in other materials.

As one notices in Fig.~\ref{fig4}, and since $\xi$ does not vary much around $T^{\textrm{bulk}}_{\textrm{g}}$ according to Eq.~(\ref{coopleng}), the crossover thickness at which the measured glass-transition temperature first shows deviations from the bulk value is a few $\xi(T^{\textrm{bulk}}_{\textrm{g}})$. Interestingly, this statement is equivalent to a purely finite-size criterion~\cite{Scheidler2000}, even though it is obtained from an explicit truncation of cooperative strings at the free interface. We may thus understand why there has been so much debate in the past between the purely finite-size effect and the mobile-layer hypothesis~\cite{Frick2000}.
\begin{figure}[b!]
\begin{center}
\includegraphics[width=8.5cm]{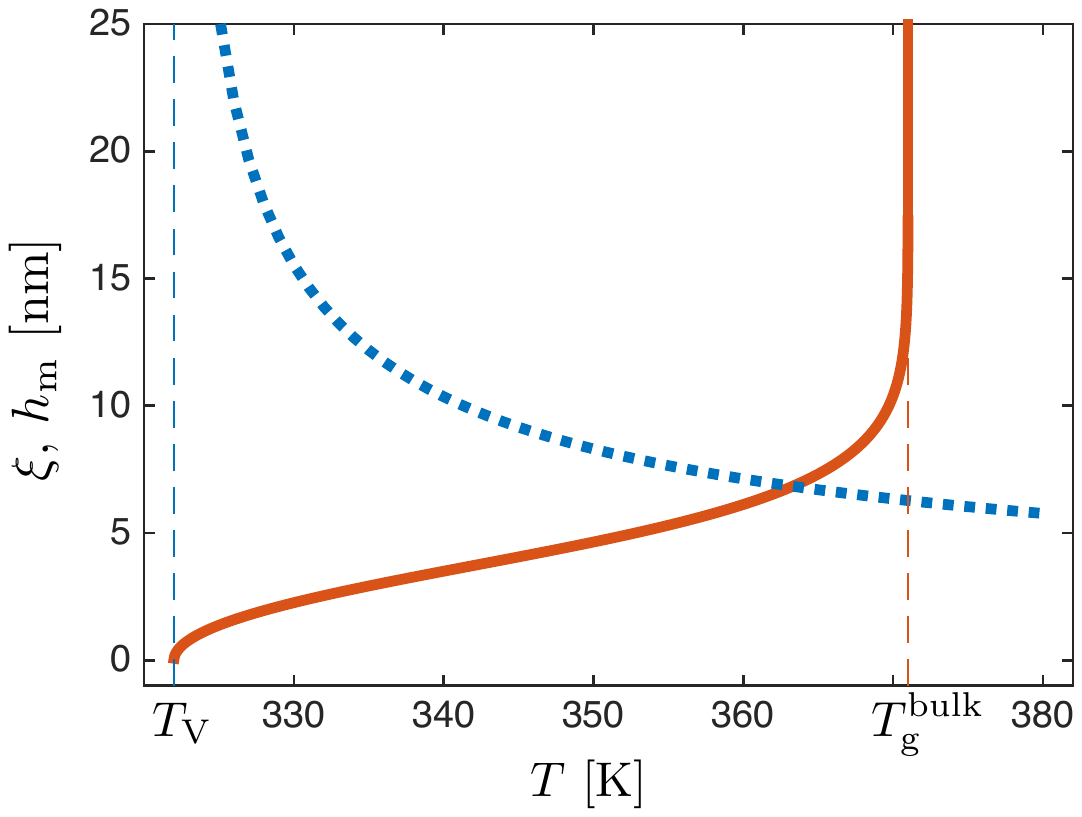}
\end{center}
\caption{\textit{Bulk cooperative length $\xi$ (dotted, Eq.~(\ref{coopleng})) and surface mobile-layer thickness $h_{\textrm{m}}$ (plain, Eq.~(\ref{mobthick})) of polystyrene, as a function of temperature. We used $T^{\textrm{bulk}}_{\textrm{g}} = 371\ \textrm{K}$~\cite{Rubinstein2003}, $T_{\textrm{c}}=463\ \textrm{K}$~\cite{Donth1996,Kahle1997}, and the values $\lambda_{\textrm{V}}\approx 3.7$~nm and $T_{\textrm{V}}\approx322\ \textrm{K}$ obtained from the fit in Fig.~\ref{fig4}.}}
\label{fig5}
\end{figure}

Since the comparison between theory and experiments (Fig.~\ref{fig4}) provides an estimate of the critical interparticle distance $\lambda_{\textrm{V}}$, we can now compute the cooperative length $\xi(T)$ for polystyrene above and below $T^{\textrm{bulk}}_{\textrm{g}}$. In Fig.~\ref{fig5}, $\xi$ increases with reducing temperature and diverges at $T_{\textrm{V}}$, as more and more molecules are required to move cooperatively for relaxation to occur. As discussed below, our model also predicts the existence of a liquid-like mobile layer at the free interface of a thin glassy film, such as that observed in~\cite{Fakhraai2008,Yang2010,Ediger2014,Chai2014,Yoon2014}. This introduces another relevant length scale to the problem: the mobile-layer thickness $h_{\textrm{m}}(T)$. An important matter of debate has been to determine the relation between $\xi$ and $h_{\textrm{m}}$, if any~\cite{Forrest2013}. In other words, is a thin-film experiment able to probe the bulk cooperative length scale $\xi$, or does it introduce another independent length scale through the mobile-layer thickness $h_{\textrm{m}}$? An advantage of the present microscopic picture is to provide a tentative answer to this question. At a given temperature $T<T^{\textrm{bulk}}_{\textrm{g}}$, there exists a position $z=h_{\textrm{m}}$ where the local relaxation time $\tau_{\textrm{s}}$ equals the bulk relaxation time at the bulk glass transition $\tau^{\textrm{bulk}}_{\textrm{g}}$. Invoking Eqs.~(\ref{wlf}) and~(\ref{main2}), this implies:
\begin{equation}
\label{mobthick}
h_{\textrm{m}}(T)=\xi(T)\ f^{\,-1}\left(\frac{T-T_{\textrm{V}}}{T^{\textrm{bulk}}_{\textrm{g}}-T_{\textrm{V}}}\right) \ .
\end{equation}
For $z>h_{\textrm{m}}$, $\tau_{\textrm{s}}$ is larger than $\tau^{\textrm{bulk}}_{\textrm{g}}$ and the system is glassy; for $z<h_{\textrm{m}}$, $\tau_{\textrm{s}}$ is lower than $\tau^{\textrm{bulk}}_{\textrm{g}}$ and the system is liquid-like. Therefore, $h_{\textrm{m}}$ is identified as the liquid-like mobile-layer thickness. As seen in Eq.~(\ref{mobthick}) and Fig.~\ref{fig5}, $h_{\textrm{m}}(T)$ is different but related to $\xi(T)$. Both lengths originate from the same cooperative-string model, but $\xi$ is a bulk quantity whereas $h_{\textrm{m}}$ reflects the truncation of the strings at the free interface. As a result, $h_{\textrm{m}}$ increases with increasing temperature and diverges at the bulk glass-transition temperature, when the entire material is in the liquid state, as observed in recent experiments~\cite{Chai2014}.

To conclude, we have developed a cooperative-string model that connects in a predictive manner essential ingredients of the glass transition, in bulk systems and near interfaces. The theory is based only on the idea of cooperativity required by increasing molecular crowding, and introduces a string-like cooperative mechanism that is motivated by recent studies. An outcome of our idealized microscopic description is to recover the Adam-Gibbs picture, as well as the Vogel-Fulcher-Tammann relation, without the need for the Doolittle ansatz to link free volume and relaxation. In particular, we derive explicit scaling expressions for the cooperativity and associated relaxation probability. Furthermore, the simplicity of the model enables application to reported anomalies in the glass transition of thin polymer films. Specifically, the free interface truncates the cooperative strings and thus enhances the mobility in its vicinity. Agreement between the present theory and reported dilatometric measurements of the glass-transition temperature of supported polystyrene films is excellent. The two adjustable parameters are: the critical interparticle distance at kinetic arrest, which is found to be similar to the persistence length of polystyrene; and the Vogel temperature which is found to be close to literature values. Finally, the model provides a way to distinguish between purely finite-size and surface effects, and to clarify the existing link between cooperative length and mobile-layer thickness. Importantly, the success of the theory applied to the thin-film data suggests that thin-film experiments are indeed relevant probes of the length scale of bulk cooperative dynamics, that may exist independently of any structural length scale in the material. This approach may be refined with additional cooperative processes, and could be adapted to the cases of attractive substrates, free-standing films, or other geometric confinements, whose effects may be crucial on the measured glass-transition temperature. 

\section*{Acknowledgments}
The authors thank Yu Chai, Olivier Dauchot, Jack Douglas, Mark Ilton, Florent Krzakala, and James Sharp for insightful discussions. This research was supported in part by NSERC of Canada and the Perimeter Institute for Theoretical Physics. Research at Perimeter Institute is supported by the Government of Canada through Industry Canada and by the Province of Ontario through the Ministry of Economic Development $\&$ Innovation.

\end{document}